\documentclass[journal=jpccck,manuscript=article,layout=twocolumn]{achemso}

\usepackage{graphicx}
\usepackage{mathrsfs}
\usepackage{hyperref}
\hypersetup{
  colorlinks = true, 	% colored Links
  allcolors = blue
}
\usepackage[utf8]{inputenc}

\usepackage{amsmath,amssymb,amstext}
\usepackage{braket}
\usepackage{bm}

\usepackage{tabularx}
\usepackage{multirow}

\newcommand{\ie}{\textit{i.e.},}
\newcommand{\eg}{\textit{e.g.},}
\newcommand{\dirimg}{./}

\SectionNumbersOn

\title{Electronic state unfolding for plane waves: energy bands, Fermi surfaces and spectral functions}

\author{David Dirnberger}
\affiliation{University of Vienna, Faculty of Physics and Center for Computational Materials Science, Vienna, Austria}
\author{Georg Kresse}
\affiliation{University of Vienna, Faculty of Physics and Center for Computational Materials Science, Vienna, Austria}
\alsoaffiliation{VASP Software GmbH, Vienna, Austria}
\author{Cesare Franchini}
\affiliation{University of Vienna, Faculty of Physics and Center for Computational Materials Science, Vienna, Austria}
\alsoaffiliation{Dipartimento di Fisica e Astronomia, Universit\`{a} di Bologna, 40127 Bologna, Italy}
\author{Michele Reticcioli}
\affiliation{University of Vienna, Faculty of Physics and Center for Computational Materials Science, Vienna, Austria}
\email{michele.reticcioli@univie.ac.at}

\begin{document}

\begin{abstract}

Modern computing facilities grant access to first-principles density-functional theory study of complex physical and chemical phenomena in materials, that require large supercell to properly model the system.
However, supercells are associated to small Brillouin zones in the reciprocal space, leading to folded electronic eigenstates that make the analysis and interpretation extremely challenging.
Various techniques have been proposed and developed in order to reconstruct the electronic band structures of super cells, unfolded into the reciprocal space of an ideal primitive cell.
Here, we propose an efficient unfolding scheme embedded directly in the Vienna \textit{Ab-initio} Simulation Package (VASP), that requires modest computational resources and allows for an automatized mapping from the reciprocal space of the supercell to primitive cell Brillouin zone.
This algorithm can compute band structures, Fermi surfaces and spectral functions, by using an integrated post-processing tool (bands4vasp).
The method is here applied to a selected variety of complex physical situations:
the effect of doping on the band dispersion in the BaFe\textsubscript{2(1-x)}Ru\textsubscript{2x}As\textsubscript{2} superconductor, the interaction between adsorbates and polaronic states on the TiO$_2$(110) surface, and the band splitting induced by non-collinear spin fluctuations in EuCd$_2$As$_2$.

\end{abstract}

\maketitle   

\section{Introduction}

Material science simulations adopting periodic boundary conditions in the framework of density functional theory (DFT) may require large unit cells in order to model long or broken periodicity in crystals.
Supercells (\ie\ large unit cells built by stacking of smaller primitive cells forming an ideal Bravais lattice), are used to study the effects of lattice impurities (\eg\ local dislocations, defects, doping), but also to investigate domain boundaries, magnetic orders, surface reactivity, and structural reconstructions, just to name a few common applications~\cite{Shanthi1998,Hine2009}.
While well developed facilities and efficient DFT packages are capable to deal with hundreds and even thousands of atoms in large cells, the analysis of the electronic properties (such as the energy band structure or the Fermi surface) gets complicated by the shrinking of the Brillouin zone (Bz) and the consequent folding of the eigenstates in the reciprocal space~\cite{Ku2010}, that prevent also a genuine comparison with photo-emission spectroscopy experiments~\cite{Day2019a,Hoekstra1987}.

\begin{figure}[ht!]
\begin{center}
\includegraphics[width=0.75\columnwidth]{\dirimg 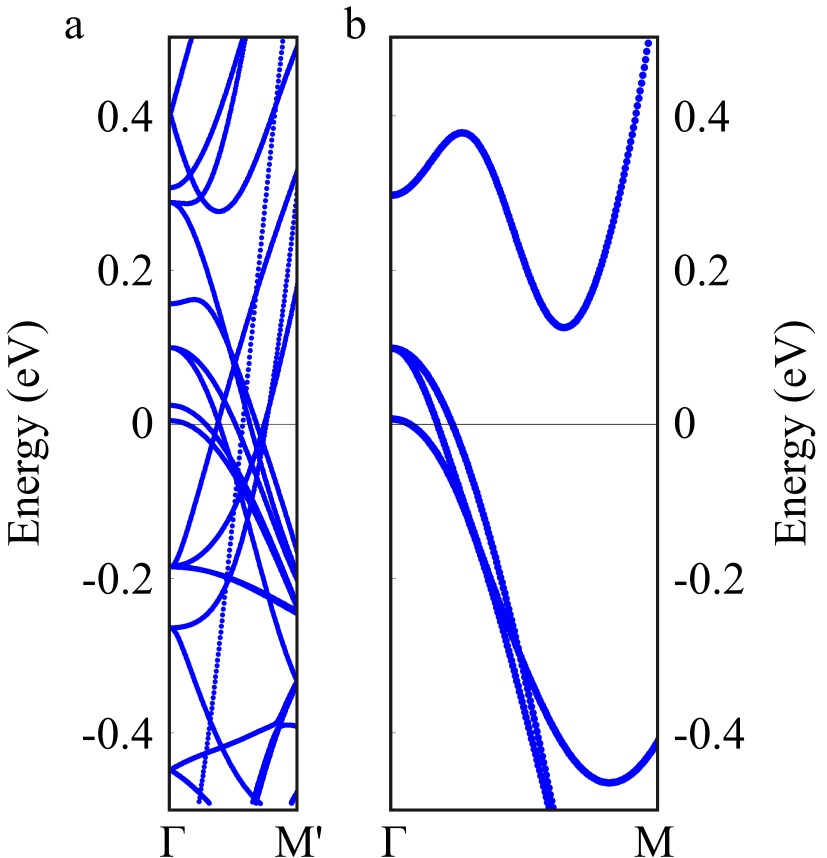}
\end{center}
\caption{Example of eigenstate folding for the pristine BaFe$_2$As$_2$ compound. Band structures obtained by using a supercell (a) and a primitive cell (b).}
\label{fig:foldexample}
\end{figure}

Figure~\ref{fig:foldexample} shows an example of the intricate band structure typically obtained by using a supercell:
Clearly, the bands calculated by using the primitive cell allow instead for a more straightforward analysis.
The intricate supercell states can be unfolded back into the larger Brillouin zone of the primitive cell by applying the unfolding technique~\cite{Ku2010, Popescu2010, Popescu2012, Boykin2007, Boykin2007a, Voit2000}.
This technique is based on the projection $P_{\bm{K}m}$ of the supercell eigenstates $\ket{\bm{K}m}$ on the primitive cell eigenstates $\ket{\bm{k}n}$:
\begin{equation} \label{blochori}
P_{\bm{K}m}(\bm{k}_i)=\sum\limits_n \vert \braket{\bm{K}m|\bm{k}_i n} \vert ^2~,
\end{equation}
where $m$ and $n$ denote energy band indices at vectors $\bm{K}$ and $\bm{k}_i$ in the reciprocal space of the supercell and primitive cell, respectively.
This projection represents the amount of Bloch character of the states $\ket{\bm{k}_i n}$ contributing to $\ket{\bm{K}m}$, which allows for a direct connection between the reciprocal space of the supercell and the primitive cell.
By weighting the contributions of all single states with $P_{\bm{K}m}$, it is indeed possible to obtain an effective band structure (EBS) of the supercell unfolded in the larger Brillouin zone of the primitive cell.

Equation~\ref{blochori} can be rewritten in terms of states of the supercell only, if the eigenstates $\ket{\bm{K}m}$ are expanded in terms of a plane-wave basis set with coefficients $C_{m,\bm{K}}$:
\begin{equation} \label{blochchar}
P_{\bm{K}m}(\bm{k}_i)=\sum_{\{\bm{g}\}} \vert C_{m,\bm{g}+\bm{k}_i} \vert ^2~.
\end{equation}
By expressing $P_{\bm{K}m}$ in this form, all information required from the primitive cell are purely geometric and collected by the reciprocal lattice vector $\bm{g}$ applied to the $\bm{k}_i$ vectors in the reciprocal space.
This alternative formulation brings the advantage to avoid any calculation on the primitive cell as well as any direct comparison between the two spaces, which could turn out to be technically challenging.

The Vienna \textit{Ab-initio} Simulation Package (VASP) is an optimal candidate for the implementation of unfolding calculations:
This code can deal with large cells efficiently, and plane waves are used as basis functions.
Moreover, a basic implementation of Eq.~\ref{blochchar} is already available in the recent VASP releases~\cite{Eckhardt2014a}.
In this work we extend the original unfolding scheme aiming to reduce the memory requirements, and to simplify the user interface for both input parameters and extraction of output data.
Specifically, we implemented an automatized scheme for generating the supercell reciprocal-space vectors $\bm{K}$ starting from given $\bm{k}_i$ vectors in the primitive cell space.
The calculation of the $P_{\bm{K}m}$ projection can be limited only to the automatically determined ($\bm{K}$, $\bm{k}_i$) pairs of interest, saving memory resources in the calculation.
The user interface has been simplified, including an automatic initialization of the primitive cell;
moreover, the user is provided with a post-processing package for the analysis of the results, ``bands4vasp''~\cite{bands4vasp}.
This updated implementation of the unfolding technique in VASP can be efficiently applied to a wide range of physical problems:
in the following we show examples of such applications, including a benchmark on the superconductivity emerging on BaFe$_2$As$_2$ upon Ru doping, the interplay between polarons and adsorbates on semiconducting TiO$_2$(110) surface, and the non-collinear magnetic ordering on EuCd$_2$As$_2$.

\section{Methodology}

\begin{figure}[ht!]
\begin{center}
\includegraphics[width=0.9\columnwidth]{\dirimg 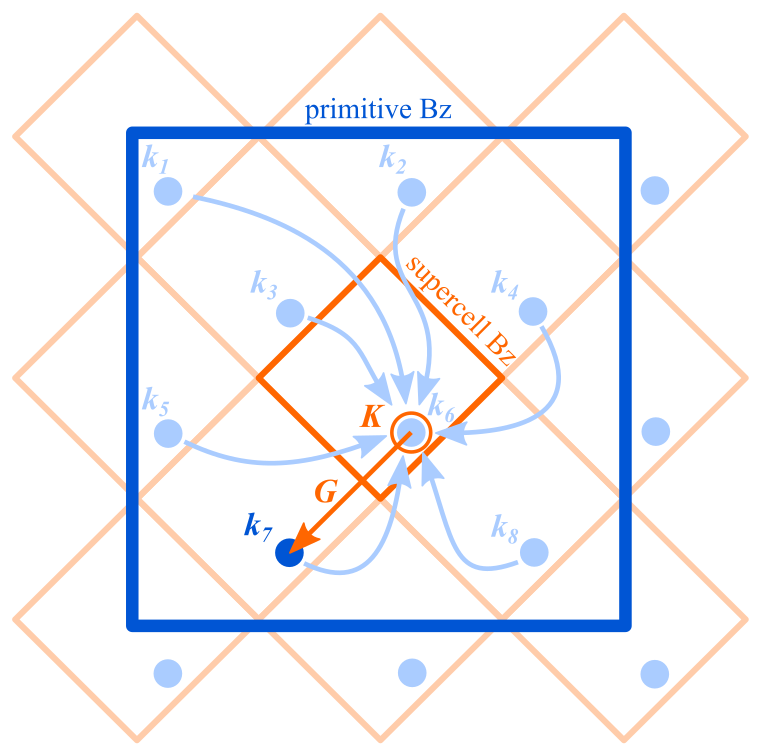}
\end{center}
\caption{Sketch of the folding problem for a bi-dimensional $2\sqrt2\times2\sqrt2$ supercell. Primitive cell Bz and supercell Bz are represented by large blue and small orange squares, respectively. The arrows indicate the folding of eigenstates from the $\lvert \bm{M}_{2\sqrt2\times2\sqrt2} \rvert=8$ different $\bm{k}_i$ primitive cell vectors (filled circles) into one $\bm{K}$ supercell vector (open, orange circle). The reciprocal lattice vector $\bm{G}$ highlights the eigenstates unfolding from the $\bm{K}$ supercell vector to a selected $\bm{k}_i$ primitive cell vector.}
\label{fig:KMk}
\end{figure}

Electronic eigenstates calculated for supercells, can be unfolded into the reciprocal space of the primitive cell by using external packages based on different methods~\cite{Zheng2015a, Medeiros2014,Medeiros2015, Herath2019, Wang2019b}, or directly in VASP that implements Eq.~\ref{blochchar}~\cite{Eckhardt2014a, Liu2016, Reticcioli2016, Reticcioli2017b}.
We have optimized this unfolding algorithm by carefully considering the relation between the $\bm{K}$ and $\bm{k}_i$ vectors, as discussed below and sketched in Figure~\ref{fig:KMk}.

The supercell and the primitive cell are described in terms of $3\times3$ matrices ($\bm{A}$ and $\bm{a}$, respectively) constructed by the corresponding lattice vectors;
similarly, the Brillouin zones are defined by matrices ($\bm{B}$ and $\bm{b}$) built by the reciprocal lattice vectors.
A transformation matrix $\bm{M}$ with integer elements relates the supercell and primitive cell, in both the direct and reciprocal spaces~\cite{Reticcioli2016}:
\begin{equation}
\begin{split}
\label{eq:Ama}
\bm{A} &=\bm{M} \bm{a} ~,\\
\bm{B} &= \left ( \bm{M}^{-1} \right )^T \bm{b} ~.
\end{split}
\end{equation}
The determinant $\lvert \bm{M} \rvert$ of the transformation matrix defines the ratio between the supercell and primitive cell volumes in the direct ($V$ and $v$) and reciprocal ($W$ and $w$) spaces: $\lvert \bm{M} \rvert = V/v = w/W$~.
The eigenstates of a single $\bm{K}$ point in the supercell Bz correspond to eigenstates of the primitive cell folded from different primitive cell points $\bm{k}_i$, with $i$ running from $1$ to $\lvert \bm{M} \rvert$ (see Fig.~\ref{fig:KMk});
these points are connected by linear combinations of supercell reciprocal lattice vectors $\{\bm{G}\}_i$:
\begin{equation}
\label{eq:kKGi}
\bm{k}_i \leftarrow \bm{K} + \{\bm{G}\}_i~\textrm{with}~i=1,\dots,\lvert \bm{M} \rvert~.
\end{equation}
The folding problem can be equivalently expressed by considering that the eigenstates of one $\bm{k}_i$ vector fold to one unique $\bm{K}$ point in the supercell first Brillouin zone, determined by one specific combination of supercell reciprocal lattice vectors $\{\bm{G}\}_0$:~\cite{Popescu2012}
\begin{equation}
\label{eq:KkG0}
\bm{K} \leftarrow \bm{k_i} - \{\bm{G}\}_0 ~,
\end{equation}
see also the straight arrow in Fig.~\ref{fig:KMk}.

The two equations above allow for an efficient mapping of supercell and primitive cell reciprocal spaces.
%In the original implementation, the user defines the supercell $\bm{K}$ vectors, and the program determines the connected primitive cell $\bm{k}_i$ vectors, in agreement with Eq.~\ref{eq:kKGi}.
%The Bloch character $P_{\bm{K}m}(\bm{k}_i)$ is then calculated for all $\bm{K}$ and $\bm{k}_i$ vectors, including ($\bm{K}$, $\bm{k}_i$) pairs not matching the two equations, that trivially result in $P_{\bm{K}m}(\bm{k}_i)=0$.
We implemented the possibility to limit the calculation of Bloch character to specific ($\bm{K}$, $\bm{k}_i$) pairs fulfilling Eq.~\ref{eq:kKGi} for any supercell $\bm{K}$ vector defined in input, excluding all other pairs which would trivially result in $P_{\bm{K}m}(\bm{k}_i)=0$.
This restriction reduces considerably the computational effort, as fewer Bloch characters need to be evaluated: the number of evaluated characters for states on any $\bm{K}$ is given by $\lvert \bm{M} \rvert$.

Additionally, 
%in the original approach one obtains the contribution from all $\bm{k}_i$ vectors to every eigenstate of any given supercell $\bm{K}$ vector, while, typically, the user is interested in the inverse problem, that is retrieving the eigenstates for selected $\bm{k}_i$ vectors from the folded supercell states.
the calculation of the Bloch character can be further limited to selected $\bm{k}_i$ vectors of interest.
In fact, the user is typically interested in retrieving the eigenstates for selected $\bm{k}_i$ vectors from the folded supercell states, rather than exploring all contributions to the supercell $\bm{K}$ vectors.
Therefore, the $\bm{k}_i$ vectors can be initialized by the user, then they get automatically translated in the supercell reciprocal space by the transformation
\begin{equation}
\label{eq:KMk}
\bm{K}^B= \bm{M} \bm{k}_i^b ~,
\end{equation}
where $\bm{K}^B$ and $\bm{k}_i^b$ represent the vector coordinates expressed in the supercell and primitive cell reciprocal spaces, respectively.
The calculation of the Bloch character can be then executed as in the original implementation, but it is limited to ($\bm{K}$, $\bm{k}_i$) pairs satisfying Eq.~\ref{eq:KkG0}:
This approach drastically reduces the computational effort of the algorithm, since the $P_{\bm{K}m}(\bm{k}_i)$ character needs to be evaluated on only one single $\bm{K}$ for any given $\bm{k}_i$ (see also discussion in the Benchmark Section).

The implementation of these automatized features simplifies the initialization of the unfolding calculation for the user.
Moreover, the primitive cell lattice vectors can also be automatically determined from the supercell by the program, by simply inverting Eq.~\ref{eq:Ama}, if the transformation matrix $\bm{M}$ is specified in input: $\bm{a} = \bm{M}^{-1} \bm{A}$.

The extraction of the output data is quite straightforward as well.
In order to further facilitate the analysis of unfolding calculations, we make available a post-processing package for band structure analysis (bands4vasp)~\cite{bands4vasp}, that can also be used for the construction of unfolded band structures, Fermi surfaces and spectral functions, as well as the automatic calculation of Fermi vectors (\ie\ the $\bm{k}_i$ vector of eigenstates at the Fermi level).
We recall that, in the framework of unfolding calculations, the spectral function $A$ is approximated as
\begin{equation}
\label{eq:specfun}
A(\bm{k},E)=\sum_m P_{\bm{K}m}(\bm{k})\delta(E_m - E)~,
\end{equation}
where $\delta(E_m - E)$ are Dirac delta functions centered around $E_m$ energies.

%%%%%%%%%%%%%%%%%%%%%%%%%%%%%%%%%%%%%%
%%%%%%%%%%%%%%%%%%%%%%%%%%%%%%%%%%%%%%
%%%%%%%%%%%%%%%%%%%%%%%%%%%%%%%%%%%%%%

\section{Benchmark and Results}

We tested our implementation of the unfolding algorithm embedded in VASP by considering the electronic properties of BaFe\textsubscript{2(1-x)}Ru\textsubscript{2x}As\textsubscript{2} (with $x=0$ and $0.25$ for the undoped and doped cases, respectively).
This material is indeed a good testbed, since many reference data are available in literature~\cite{Reticcioli2016,Reticcioli2017b, Wang2013}.
We take this opportunity also to describe features included in the post-processing bands4vasp package, such as the visualization of band structures, Fermi surfaces and spectral function, and the calculation of Fermi vectors (Sec.~\ref{sec:bafe2as2}).
Finally, we show novel and more challenging applications of our machinery:
by performing calculations on large cells modeling the TiO$_2$(110) surface, the unfolding analysis reveals formation of flat bands originating from in-gap polaronic states that are perturbed by the interaction with CO adsorbates deposited on the material surface (Sec.~\ref{sec:tio2}).
Moreover, we describe the band splitting occurring in the transition process from non-collinear paramagnetic to ferromagnetic ordering in EuCd$_2$As$_2$ (Sec.~\ref{sec:eu}).

%%%%%%%%%%%%%%%%%%%%%%%%%%%%%%%%%%%%%%
%%%%%%%%%%%%%%%%%%%%%%%%%%%%%%%%%%%%%%
%%%%%%%%%%%%%%%%%%%%%%%%%%%%%%%%%%%%%%

\subsection{Analysis of metallic states in BaFe\textsubscript{2(1-x)}Ru\textsubscript{2x}As\textsubscript{2}}
\label{sec:bafe2as2}

The metal-to-superconductor transition driven by Ru doping in the BaFe$_2$As$_2$ pnictide has attracted wide interest in both theoretical and experimental communities~\cite{Thaler2010, Sharma2010, Eom2012, Kim2013, Sharma2013, Devidas2014, Sen2014a, Liu2015}, and the unfolding technique has proven itself largely useful to support density-functional theory investigations:
One of the most evident results observed from the effective band structures is the progressively closure of hole pockets upon doping, due to a coupling with structural distortions~\cite{Wang2013,Reticcioli2017b}.
%Here, we show that our implementation of the unfolding algorithm implemented in VASP is able to correctly reproduce these results, at a smaller computational cost than the original version.

We performed spin-unpolarized DFT calculations on BaFe$_2$As$_2$ by maintaining a similar computational setup as in Ref.~\citenum{Reticcioli2016}, but applying the updated version of the unfolding algorithm.
We modeled the BaFe$_2$As$_2$ structure adopting supercells of different size ($\bm{A}_2$,$\bm{A}_8$,$\bm{A}_{16}$), constructed by applying the following transformation matrices (as described in Eq.~\ref{eq:Ama})
\begin{equation*}
\begin{split}
\bm{M}_{2,8,16}=&\left( \begin{array}{rrr}
0 & 1 & 1 \\
1 & 0 & 1 \\
1 & 1 & 0 \\
\end{array} \right)
,\left( \begin{array}{rrr}
0 & 2 & 2 \\
2 & 0 & 2 \\
1 & 1 & 0 \\
\end{array} \right)
,\\
\textrm{and}& \left( \begin{array}{rrr}
0 & 4 & 4 \\
2 & 0 & 2 \\
1 & 1 & 0 \\
\end{array} \right)
,
\end{split}
\end{equation*}
with determinants $\lvert \bm{M} \rvert=2,8,16$, respectively.
The $\bm{M}_2$ matrix represents the transformation from primitive cell to conventional unit cell for compounds with the body-centered tetragonal I4/mmm space group such as BaFe$_2$As$_2$~\cite{Setyawan2010}:
We remark that the $\bm{M}$ matrix elements are required to be integer in the unfolding formalism, but this requirement does not prevent us to model rotations or non trivial transformations~\cite{Boykin2007a}.

\begin{figure}[ht!]
\begin{center}
\includegraphics[width=1.0\columnwidth]{\dirimg 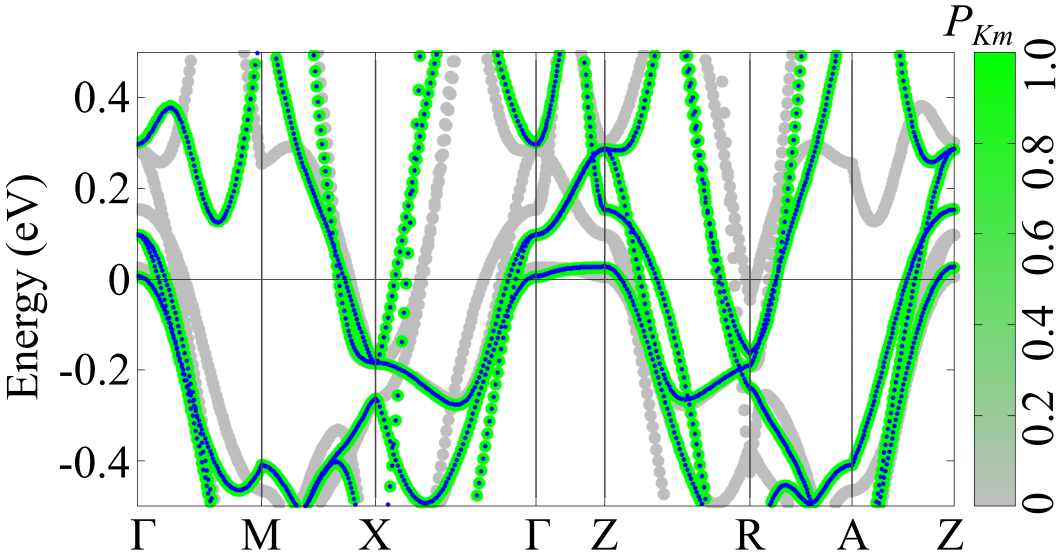}
\end{center}
\caption{Effective band structure of the supercell constructed by the $\bm{M}_2$ transformation matrix (gray-to-green color gradient), compared to the band structure of primitive cell (blue).}
\label{fig:ba1}
\end{figure}

We performed a preliminary test by considering the pure (undoped) BaFe$_2$As$_2$ crystal and comparing the effective band structure obtained from the supercell with the bands calculated directly for the primitive cell.
In fact, no difference should be observed between the two band structures, once the supercell states are unfolded into the reciprocal space of the primitive cell.
Figure~\ref{fig:ba1} shows the unfolded bands of the supercell $\bm{A}_2$ constructed by the transformation matrix $\bm{M}_2$.
As expected (due to $\lvert \bm{M}_2 \rvert=2$), we obtained for the supercell double as many of the bands obtained directly from the primitive cell calculation, shown in the figure for comparison.
The unfolding algorithm is able to correctly identify the band folding, and to assign the Bloch character $P_{\bm{K}m}$ accordingly:
the gray color in the gradient palette identifies the folded bands with $P_{\bm{K}m}=0$ that belong to different points in the primitive cell reciprocal space.
Usually, the folded bands should not be considered in the analysis of the electronic properties of the material, and can be omitted from the graph (\eg\ by setting the gradient palette with $P_{\bm{K}m}=0$ to the same color as the background of the image).

\begin{figure*}[ht!]
\begin{center}
\includegraphics[width=1.6\columnwidth]{\dirimg 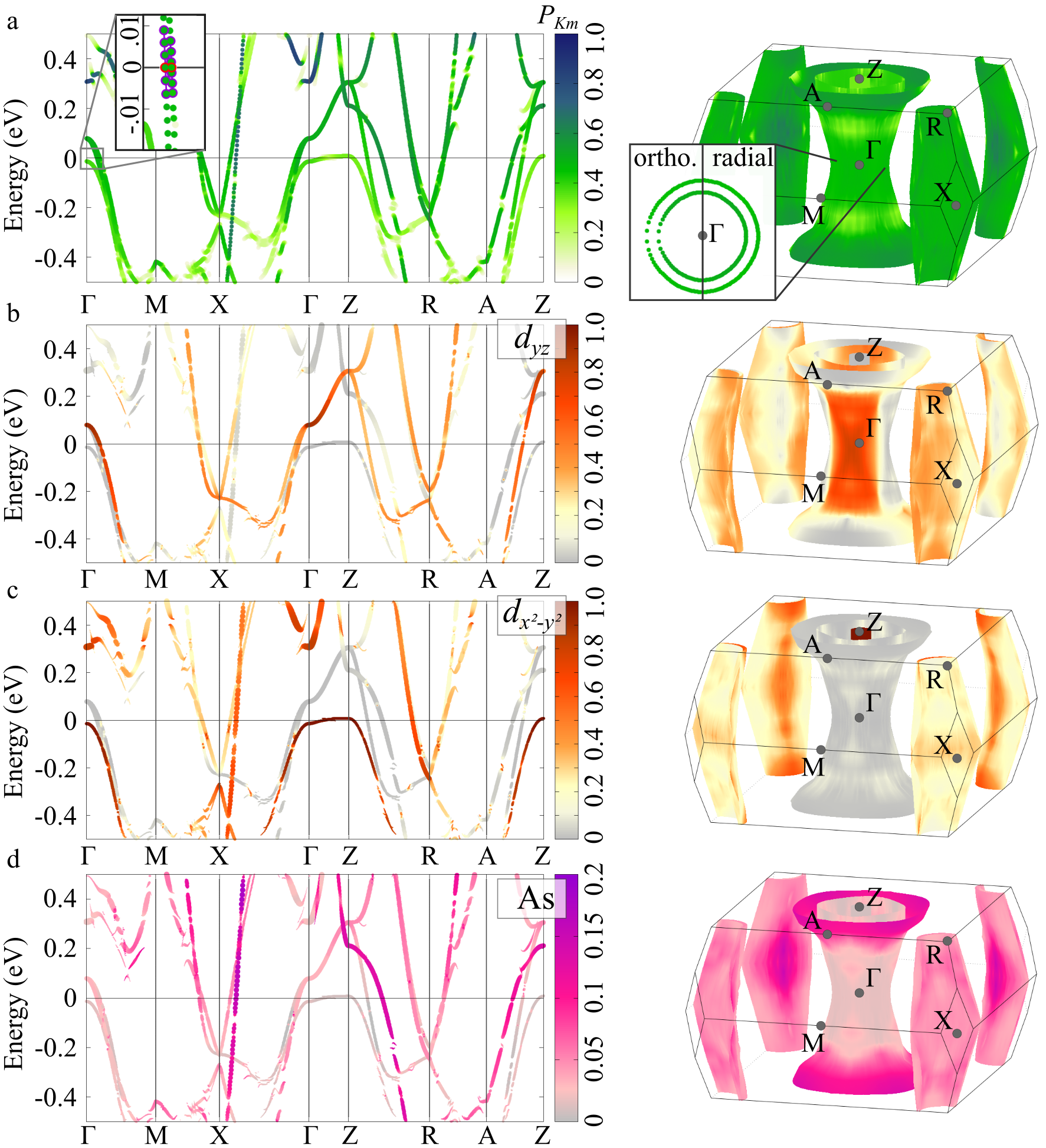}
\end{center}
\caption{Unfolding for the doped BaFe\textsubscript{2(1-x)}Ru\textsubscript{2x}As\textsubscript{2} ($x=0.25$). a: effective band structure and 3-dimensional Fermi surface; the color gradient indicates the Bloch character; the insets show the identification of the Fermi vectors, and a 2-dimensional cut of the Fermi surface with orthogonal (left) and radial (right) sampling of the reciprocal space. b--d: band structures with the Bloch character represented by variable point size, while the color gradient indicates the $d_{xz}$ (b) and $d_{x^2-y^2}$ (c) orbital character of Fe atoms, and the overall contribution from As atoms (d); the corresponding 3-dimensional Fermi surfaces are also shown.}
\label{fig:badop}
\end{figure*}

Figure~\ref{fig:badop} shows the effects of $25\%$ of Ru doping on the electronic properties of BaFe\textsubscript{2(1-x)}Ru\textsubscript{2x}As\textsubscript{2} ($x=0.25$).
The band structure in Fig.~\ref{fig:badop}a correctly reproduces the closure of the hole pockets around the $\Gamma$ and Z high symmetry points~\cite{Reticcioli2017b}.
The intersection of the bands with the Fermi level can be identified automatically, by using the bands4vasp tool (see inset in Fig.~\ref{fig:badop}a):
the eigenstates around Fermi are assigned to different bands by considering the Bloch character and the orbital symmetry, and the intersection of every band with Fermi is found by an interpolation.
This feature is extremely useful for the analysis of the Fermi vectors, that get progressively shortened for the hole pockets around $\Gamma$ upon Ru doping in this material.
Moreover, by collecting all Fermi vectors in the Brillouin zone, it is possible to construct 3-dimensional Fermi surfaces (see right panel in Fig.~\ref{fig:badop}a).
The 2-dimensional cut of the basal plane around $\Gamma$ (see inset in Fig.~\ref{fig:badop}a) highlights the importance of an accurate sampling of the reciprocal space, by comparing the Fermi surface obtained using two different approaches:
The right side of the Fermi surface is constructed by using a radial distribution of the $k$ points, leading to a better resolved description of the states around $\Gamma$, as compared to the resolution obtained by adopting a conventional rectangular grid (left side).

The bands4vasp tool can also efficiently extract the atomic orbital character of the eigenstates from unfolding calculations.
Figures~\ref{fig:badop}b--d show the effective band structures with Bloch character represented by the size of the circles, and the color gradient representing the projection of the states on the $d_{yz}$ (Fig.~\ref{fig:badop}b) and $d_{x^2-y^2}$ (Fig.~\ref{fig:badop}c) orbitals (essentially due to Fe atoms), and the overall contribution from As atoms (Fig.~\ref{fig:badop}d).
The contribution of Fe states around $\Gamma$ ($d_{yz}$, and $d_{xz}$ not shown here), Z ($d_{x^2-y^2}$) and on the Bz border stems out clearly from both the band structure and the corresponding 3-dimensional Fermi surfaces.
Similarly, the hybridization with As atoms leads to states at the Fermi level around Z and $X$.

\begin{figure*}[ht!]
\begin{center}
\includegraphics[width=1.4\columnwidth]{\dirimg 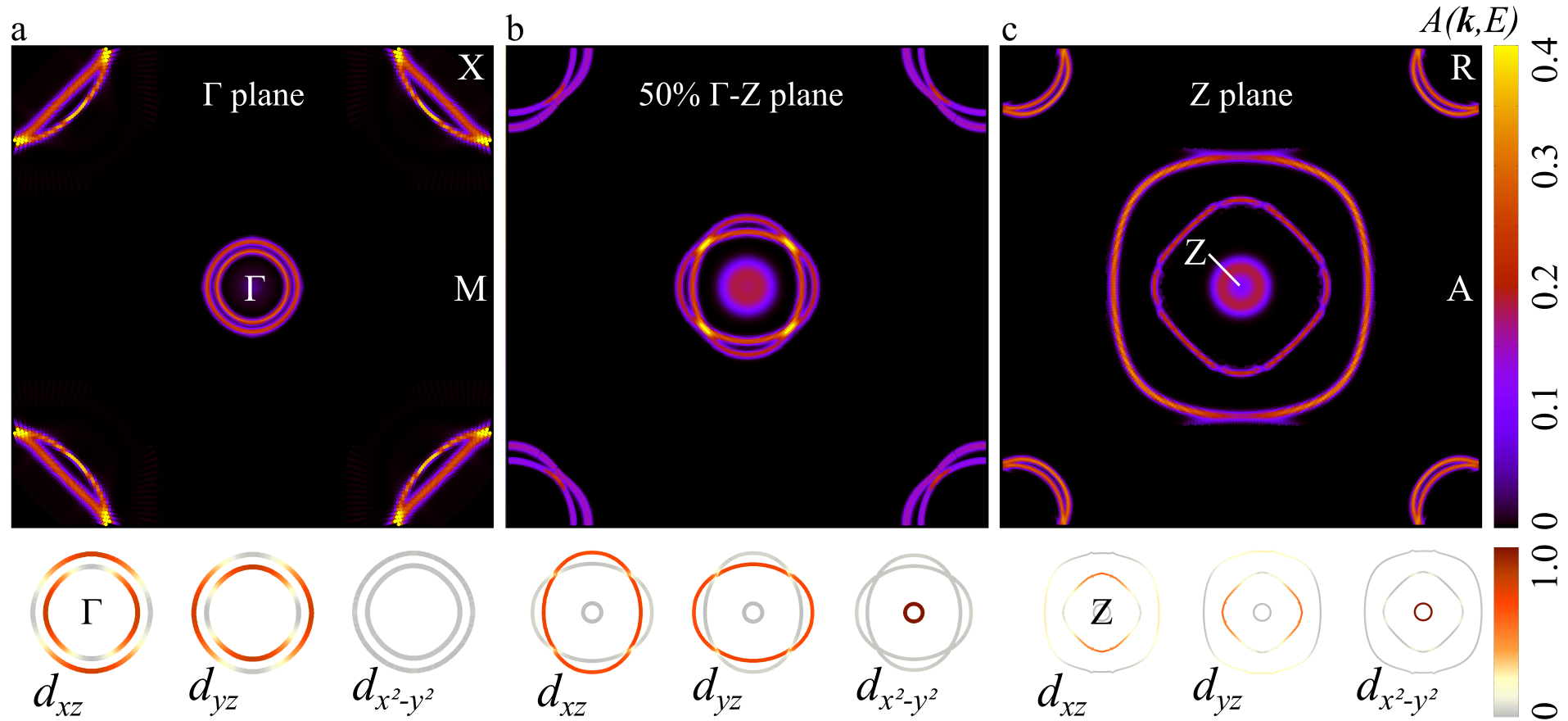}
\end{center}
\caption{2-dimensional Fermi surfaces calculated using the spectral function $A(\bm{k},E_{\rm Fermi})$ for the basal plane (a), and the parallel planes half way in the $\Gamma$--Z direction (b) and including the high-symmetry point Z (c). The corresponding orbital characters of the central rings are shown in the bottom of each panel.}
\label{fig:spec}
\end{figure*}

By looking carefully at the $d_{x^2-y^2}$ orbital (in Fig.~\ref{fig:badop}c), we note a band crossing the Fermi level in the $\Gamma$--Z direction.
This feature is highlighted in Figure~\ref{fig:spec}, that shows 2-dimensional Fermi surfaces for the basal plane and the parallel planes along $\Gamma$--Z, obtained for the spectral function $A(\bm{k},E_{\rm Fermi})$.
On the basal plane, only two states appear sharply around $\Gamma$ ($d_{xz}$ and $d_{yz}$, see orbital analysis on the bottom images in Fig.~\ref{fig:spec}a).
Conversely, the spectral function for the inner state (with $d_{x^2-y^2}$ orbital symmetry) is absent, and becomes progressively better defined when moving towards Z (Fig.~\ref{fig:spec}b,c).
The spectral function allows us also to easily identify band degeneracy and crossing points between different bands, that are revealed by a more intense value:
In Fig.~\ref{fig:spec}b the crossing of the $d_{xz}$ and $d_{yz}$ states determines four points with high value for the spectral function around the center of the plane.
This crossing corresponds to a progressive band switching between $d_{xz}$ and $d_{yz}$ states, clearly identifiable by looking at the evolution of the orbital symmetry of the internal and external rings moving from $\Gamma$ to Z.

We conclude our benchmark by commenting on memory requirements of the unfolding algorithm.
The automatic determination of ($\bm{K}$, $\bm{k}_i$) pairs (as in Eqs.~\ref{eq:KkG0} and~\ref{eq:KMk}) reduces the computational effort, by limiting the unfolding procedure only to points of interest.
%Table~\ref{tab} reports the ratio between the amount of memory used by the newer implementation of the unfolding algorithm, and the corresponding calculation performed via the previous version.
For small systems, as the $\bm{A}_2$ supercell, we counted a memory gain of about 20\% when using 300 $\bm{k}$ points, which increases to approximately 30\% and 50\% for the larger $\bm{A}_8$ and $\bm{A}_{16}$ cells, respectively.
The lower computational requirements are very useful for performing unfolding calculations that model complex systems:
some original application of the unfolding method are presented in the following sections.

%%%%%%%%%%%%%%%%%%%%%%%%%%%%%%%%%%%%%%
%%%%%%%%%%%%%%%%%%%%%%%%%%%%%%%%%%%%%%
%%%%%%%%%%%%%%%%%%%%%%%%%%%%%%%%%%%%%%

\subsection{In-gap polaronic states and adsorbates on TiO$_2$(110) surface}
\label{sec:tio2}

\begin{figure*}[ht!]
\begin{center}
\includegraphics[width=2\columnwidth]{\dirimg 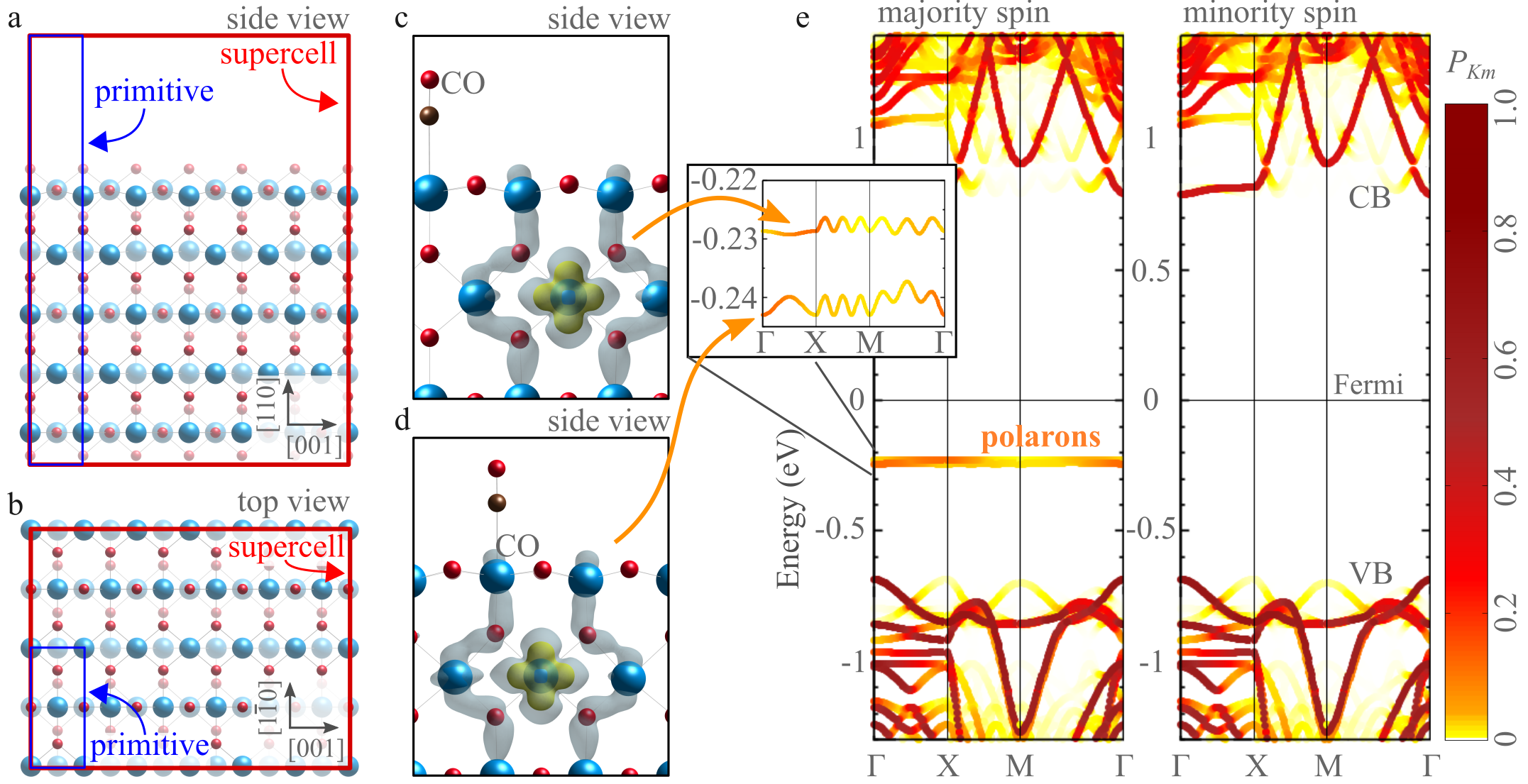}
\end{center}
\caption{Unfolding for surface slab calculations. Side (a) and top (b) view of the pristine rutile TiO$_2$(110) surface; the blue and red rectangles indicate the primitive cell and the $6\times2$ super cell, respectively. Panels c,d: Detail of the two polarons and the two adsorbed CO molecules; the gray and yellow areas represent the polaronic charge at different isosurface levels. Panel e: corresponding effective band structure; the inset shows a detail of the polaronic bands, between the conduction (CB) and valence (VB) bands.}
\label{fig:tio2}
\end{figure*}

The unfolding algorithm can be applied also for DFT calculations on systems with reduced dimensionality, such as the surface slab shown in Figure~\ref{fig:tio2}, modeling the pristine rutile TiO$_2$(110) termination.
Unit cells modeling surfaces of solids in VASP contain a vacuum region in order to interrupt the periodicity of the system along the surface normal~\cite{Sholl2009}.
Typically, several atomic layers are also included in the model in order to mimic the properties of the bulk below the surface (see Fig.~\ref{fig:tio2}a).
Primitive cells and supercells share the same vector perpendicular to the surface as in Fig.~\ref{fig:tio2}a;
in order to model surface reconstructions or defects, supercells are constructed by enlarging the lateral size (see Fig.~\ref{fig:tio2}b), leading to folding of electronic states in analogy with the bulk.

Rutile TiO$_2$(110) supercells can be used to model the formation of oxygen vacancies on the surface, that lead to stabilization of small electron polarons, \ie\ electrons strongly localized on Ti ions and coupled with the phonon field~\cite{Setvin2014,Rousseau2020,Franchini2021}.
Small polarons are typically associated to eigenstates appearing in the energy band gap of semiconductors;
moreover, polaron localization can occur on different sites, with different formation energy (in rutile, sub-surface Ti ions are preferred, over surface sites)~\cite{Reticcioli2018a}.
Polarons are known to drastically affect the electronic and chemical properties of the hosting material, with substantial impact on the applications:
we focus here on the chemical activity of rutile surface, by considering the interplay between polarons and CO adsorbates, and the effects on the eigenstates~\cite{Reticcioli2019c}.

Figures~\ref{fig:tio2}c--e collect the results obtained for large $6\times 2$ supercells containing 363 atoms, including two CO molecules and two polarons (technical details of the calculation in Ref.~\citenum{Reticcioli2019c}).
As shown in Figs.~\ref{fig:tio2}c,d, the CO can adsorb on Ti sites at different distance from the polarons.
The corresponding effective band structure (unfolded on the surface primitive cell) shows the appearance of the strongly localized polaronic states, revealed by two flat in-gap bands (one per polaron) in the majority spin channel (Fig.~\ref{fig:tio2}e).
By looking closer at these in-gap bands (inset in Fig.~\ref{fig:tio2}e), we note that the two polaronic states are not degenerate, due to the interaction with the CO molecules.
The band appears more perturbed for the polaronic state closer to the CO molecule, as manifested by the increased band width.
Perturbations of the polaronic in-gap states may originate also from the repulsive interaction of polarons at small distance, as described in Ref.~\citenum{Reticcioli2021} for polarons and bipolarons.

%/home/lv70338/mr_oxsur/TiO2/S0S1-COatS0-COnopol

%%%%%%%%%%%%%%%%%%%%%%%%%%%%%%%%%%%%%%
%%%%%%%%%%%%%%%%%%%%%%%%%%%%%%%%%%%%%%
%%%%%%%%%%%%%%%%%%%%%%%%%%%%%%%%%%%%%%

\subsection{Non-collinear ferromagnetic fluctuations in EuCd$_2$As$_2$}
\label{sec:eu}

We describe here the application of the unfolding algorithm on systems with non-collinear magnetic ordering~\cite{Medeiros2015}.
We consider the paramagnetic-to-ferromagnetic transition in EuCd$_2$As$_2$, an interesting semimetal showing emergence of Weyl fermions in the paramagnetic phase due to spin fluctuations of Eu magnetic moments~\cite{Maeaaw4718}.

\begin{figure}[ht!]
\begin{center}
\includegraphics[width=1\columnwidth]{\dirimg 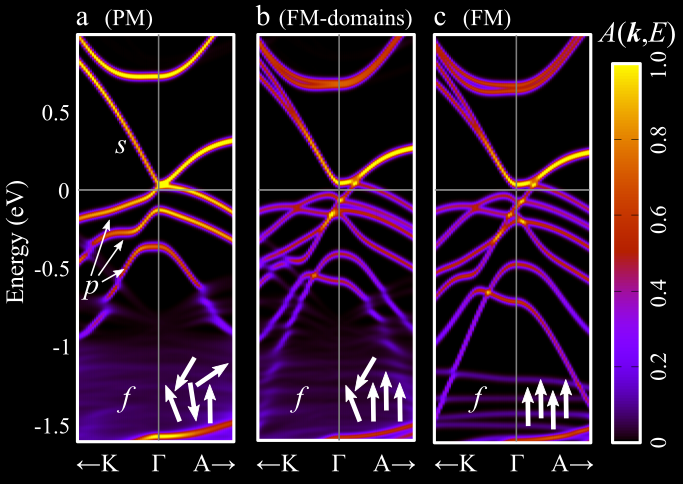}
\end{center}
\caption{Effective band structure obtained by non-collinear-spin calculations for EuCd$_2$As$_2$ considering a paramagnetic phase (a), ferromagnetic domains (b) and perfect ferromagnetic ordering (c).
The band structures focus around the $\Gamma$ point from the $K-\Gamma-A$ path. The white arrows sketch the arrangement of the magnetic moments of the Eu atoms.}
\label{fig:eu}
\end{figure}

Figure~\ref{fig:eu} compares the effective spectral functions calculated for EuCd$_2$As$_2$ with different magnetic orderings (technical details of calculations are described in Ref.~\citenum{Maeaaw4718}).
The paramagnetic phase was modeled by a large supercell including 16 Eu atoms with magnetic moments fixed to random orientations, resulting in a vanishing total magnetization:
the corresponding spectral function unfolded in the reciprocal space of the primitive cell is shown in Fig.~\ref{fig:eu}a.
The flat $f$ bands of Eu atoms appearing around $-1.5$~eV show an evident incoherence, due to the random orientation of the magnetic moments.
The three $p$ bands of As atoms appear strongly spin-degenerate:
the spectral function is very effective in capturing the band degeneracy, as degenerate bands result in higher values of the spectral character integrating the contribution from every state, as described in Eq.~\ref{eq:specfun} (at variance with band structures, showing instead the Bloch character of every state individually).

The ferromagnetic phase shows interesting changes (Fig.~\ref{fig:eu}c).
First, we note that the $f$ bands are more coherent, as expected, due to the ferromagnetic alignment of all Eu magnetic moments.
Remarkably, the ferromagnetic order induces a splitting of the $p$ states of As atoms:
we observe indeed six bands, lifting the spin degeneracy of the three $p$ bands in the paramagnetic phase (note also the lower spectral function value, as compared to the paramagnetic case).

Although the study of ferromagnetic systems could be done directly in the primitive cell, the supercell approach allowed us to study the paramagnetic-to-ferromagnetic transition by considering ferromagnetic domains embedded in a paramagnetic environment.
In Fig.~\ref{fig:eu}b, we show the spectral function of the system including a large ferromagnetic domain (consisting of 10 Eu atoms with aligned magnetic moments), and a smaller region (6 atoms) with Eu magnetic moments constrained to random directions.
The splitting of $p$ orbital persists in this transition state:
This is an example of the effect of spin fluctuations on the paramagnetic phase of the compounds.
In smaller ferromagnetic domains, the band splitting is gradually reduced, progressively converging towards the paramagnetic degeneracy (results obtained by modeling different size of the ferromagnetic domain are available in Ref.~\citenum{Maeaaw4718}).

%/home/michele/Dropbox/WORK/projects/EuCd2As2/img/fancy4website
%/data/michele/EuCd2As2/supercell/4x4x1/FM-mostly/hq-EBS

\section{Conclusions}

In summary, we report here our optimization of the unfolding scheme embedded in VASP, characterized by a simplified user interface, and reduced memory requirements, thanks to an efficient mapping between the reciprocal spaces of the supercell and primitive cells.
The construction of effective band structures, spectral functions, Fermi surfaces and projections of electronic states on orbitals and ions, is further facilitated by the bands4vasp post-processing package.

The unfolding scheme is extremely useful in the interpretation of the results obtained by supercell approaches, and it facilitates the comparison with the experimental observations, especially in the field of spectroscopy.
The application range is very broad:
We take here the BaFe\textsubscript{2(1-x)}Ru\textsubscript{2x}As\textsubscript{2} superconductor as a benchmark, given the large amount of data available in literature.
Moreover, we considered the adsorption of CO molecules on the rutile TiO$_2$(110) surface, in order to show the suitability of the algorithm for very large supercells, such as those required in surface science calculations.
In this case study, the effective band structures highlight the interactions of adsorbates with strongly localized polarons, revealed by perturbation of the flat polaronic bands.
Finally, we performed non-collinear calculations for the EuCd$_2$As$_2$ semimetal:
The supercell approach allowed us to model the paramagnetic phase, by constraining magnetic moments along random directions, resulting in a vanishing total magnetization.
The corresponding spectral function reveals the spin degeneracy of shallow states below Fermi, that is lifted by including spin-fluctuations via formation of ferromagnetic domains.

The unfolding algorithm proposed here represent a useful computational tool for a wide range of physical and chemical phenomena requiring very large supercells, thanks to the improved interface, the reduced computational requirements and the powerful analysis tool.

\section*{Acknowledgments}

This project has benefited from intensive discussions with D.D. Sarma during his visiting professorship in Vienna in spring 2018, during which he elaborated on the concept of electronic state unfolding in a much-appreciated lecture.
This work was supported by the joint FWF and Indian Department of Science and Technology (DST) project INDOX (Project No. I1490-N19).
The computational results presented have been achieved using the Vienna Scientific Cluster (VSC).

\bibliography{bib-exp}

\end{document}